\documentclass[aps,pra,10pt,twocolumn,floatfix,superscriptaddress,showpacs,numerical,footinbib]{revtex4-1}

\usepackage{graphicx}
\usepackage{amsmath, amssymb, braket}
\usepackage[T1]{fontenc}
\usepackage[utf8]{inputenc}
\usepackage[colorlinks=true,linkcolor=blue,citecolor=blue,breaklinks]{hyperref}
\usepackage[pdftex]{color}

\newcommand{\phd}{{\protect\vphantom{\dag}}}		

\definecolor{purple}{rgb}{0.5,0,0.5}

\begin{document}
\title{Phase transitions and adiabatic preparation of a fractional Chern insulator \\ in a boson cold atom model}
\author{Johannes Motruk}
\affiliation{\mbox{Department of Physics, University of California, Berkeley, California 94720, USA} and \\ \mbox{Materials Science Division, Lawrence Berkeley National Laboratory, Berkeley, California 94720, USA}}
\affiliation{\mbox{Max-Planck-Institut f\"ur Physik komplexer Systeme, N\"othnitzer Str.\ 38, 01187 Dresden, Germany}}
%
\author{Frank Pollmann}
\affiliation{\mbox{Technische Universität München, Physics Department T42, 85747 Garching, Germany}}
\affiliation{\mbox{Max-Planck-Institut f\"ur Physik komplexer Systeme, N\"othnitzer Str.\ 38, 01187 Dresden, Germany}}
\date{\today}
\begin{abstract}
We investigate the fate of hardcore bosons in a Harper-Hofstadter model which was experimentally realized by Aidelsburger \textit{et al.} [Nature Physics \textbf{11} , 162 (2015)] at half filling of the lowest band. We discuss the stability of an emergent fractional Chern insulator (FCI) state in a finite region of the phase diagram that is separated from a superfluid state by a first-order transition when tuning the band topology following the protocol used in the experiment. Since crossing a first-order transition is unfavorable for adiabatically preparing the FCI state, we extend the model to stabilize a featureless insulating state. The transition between this phase and the topological state proves to be continuous, providing a path in parameter space along which an FCI state could be adiabatically prepared. To further corroborate this statement, we perform time-dependent DMRG calculations which demonstrate that the FCI state may indeed be reached by adiabatically tuning a simple product state.
\end{abstract}
\maketitle

\section{Introduction}

Interacting particles occupying topologically non-trivial band structures have been a topic of high interest in recent years due to the exotic many-body states they can form.
Such exotic topological phases can be realized as lattice analogs~\cite{Kol1993,Moeller2009,Scaffidi2012,Moeller2015} of fractional quantum Hall (FQH) states~\cite{Tsui1982,Laughlin1983}, now understood to be first examples of a wider class of fractional Chern insulators (FCIs) in general topological band structures \cite{Tang2011,Sun2011,Sheng2011,Neupert2011a,Regnault2011,Bergholtz2013,Moeller2009,Moeller2015}.
Despite being very well characterized theoretically, FCIs have not yet been realized experimentally. Ultracold atoms in optical lattices are among the most promising candidate systems to detect these topologically ordered states~\cite{Sorensen2005,Hafezi2007,Palmer2006,Palmer2008}. As a more concrete direction, two experimental groups~\cite{Miyake2013,Aidelsburger2013} have implemented an optical lattice setup in which bosons are governed by the Harper-Hofstadter Hamiltonian~\cite{Harper1955,Hofstadter1976}, which features a considerably flat lowest band with nonzero Chern number, favoring the occurrence of an FCI state. While the aforementioned experiments focused on single-particle properties, the study of interacting particles in ladder systems described by this Hamiltonian has been reported~\cite{Tai2017} as a first step to observe many-body physics in these setups. 

The preparation of exotic many-body states in these systems poses two main challenges. Firstly, in order to engineer complex hopping matrix elements in the optical lattices, the system has to be subject to a constant periodic drive~\cite{Jaksch2003,Mueller2004,Dalibard2011}. An expansion of the periodically time-dependent Hamiltonian in the inverse driving frequency~\cite{Rahav2003,Goldman2014,Eckardt2015} leads to an effective time-independent Floquet Hamiltonian that has the desired topologically non-trivial properties~\cite{Oka2009,Lindner2011,Grushin2014,Bukov2015,Eckardt2017}. This constant driving will, at some time scale, lead to energy absorption which would destroy a topological state and in the long term drive the system to a featureless infinite temperature state~\cite{Dalessio2014,Lazarides2014}. To remedy this problem, it has been argued that this behavior may be preceded by a prethermalization region during which the system is indeed governed by an effective time-independent Hamiltonian and which extends to a time that is exponentially long in the driving frequency~\cite{Abanin2015a,Abanin2015,Kuwahara2016,Mori2016,Bukov2016,Abanin2017,Else2017}.
The second challenge is that ultracold atoms in an optical lattice typically represent an isolated quantum system which prevents the system from being cooled down in contact with an external bath to reach its ground state. One way to eliminate this problem is a (quasi-)adiabatic preparation scheme~\cite{Popp2004,Sorensen2010,He2017}. It consists of a protocol by which a state of the system that may be more easily prepared (e.g. a condensate) is guided into the final topologically ordered state by changing the properties of the optical lattice, i.e. tuning parameters in the Hamiltonian describing the system.
In order to ensure that the final state after this (quasi-)adiabatic evolution is indeed the ground state of the final Hamiltonian, it is favorable that any phase transitions crossed during the evolution are continuous~\cite{Schuetzhold2006,Barkeshli2015}. At a first-order transition, the system might otherwise stay in a metastable initial state impeding the evolution into the topologically ordered ground state.

In this work, we focus on the second challenge, i.e., if it is possible to find a path in parameter space that leads from a realistically preparable, trivial state into an FCI state while only crossing continuous phase transitions. We study a model Hamiltonian at $1/2$ filling of the lowest band motivated by the optical lattice setup of Aidelsburger \textit{et al.}~reported in~\cite{Aidelsburger2015}. While we find a first-order transition into the FCI state when increasing a parameter from the original work~\cite{Aidelsburger2015} that turns the band structure trivial if sufficiently large, we propose  an additional parameter in the model which allows reaching the FCI phase upon crossing a continuous transition. This modification introduces an additional chemical potential in form of a superlattice. We perform explicit time-dependent simulations for the adiabatic tuning of the system from a trivial into the FCI state where we show that the preparation scheme works for a path crossing the continuous transition while it is not possible when traversing the first-order transition. To study the model, we rely on density matrix renormalization group (DMRG) computations on infinite cylinders of finite circumference~\cite{White1992,McCulloch2008,Zaletel2015}.

This paper is organized as follows. We introduce the model and employed method in Sec.~\ref{sec:model} and demonstrate the appearance of an FCI state for the Hamiltonian in Sec.~\ref{sec:FCI}. In Sec.~\ref{sec:PD}, we study the phase diagram of the model and present the time-dependent calculations in Sec.~\ref{sec:time}.

\section{Model and method \label{sec:model}}

\begin{figure}[t]
 \includegraphics[width=8cm]{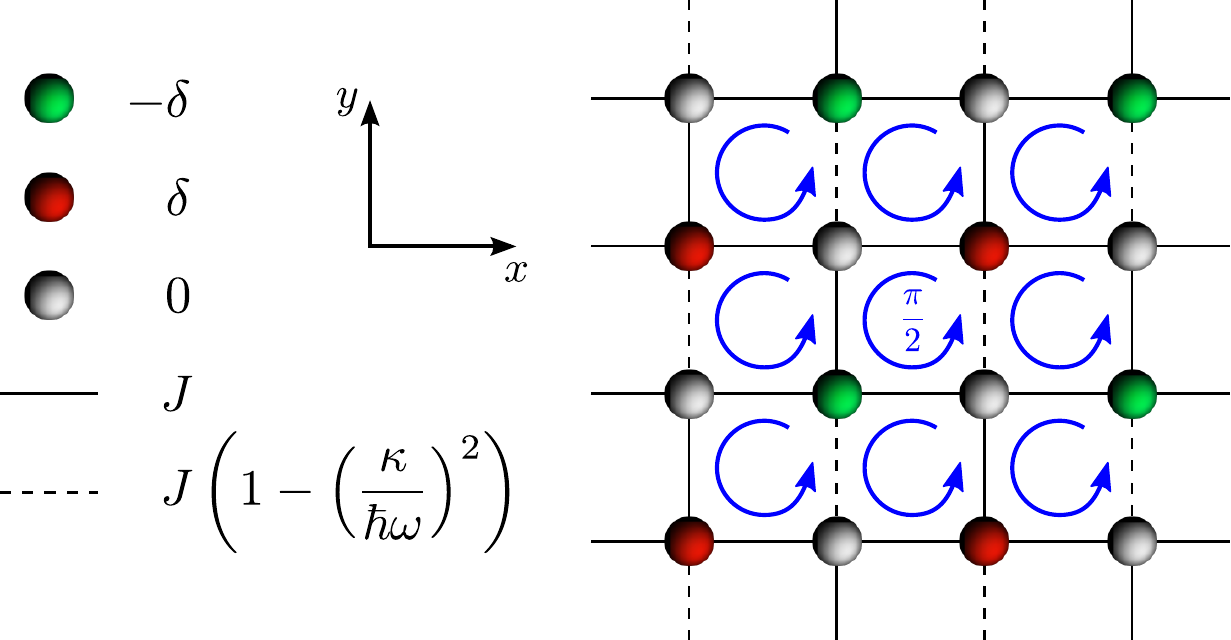}
 \caption{Harper-Hofstadter model considered in this study. The model has a four site unit cell containing sites with energy offset $-\delta, 0$ and $\delta$ and the hopping in {$y$-direction} is renormalized according to $J(1+f_{x,y})$ \eqref{eq:fmn}. We chose $\phi_0 = 0$ to obtain the hopping amplitude in this illustration as used throughout this work. We also omit the phases of the hopping in this depiction, each square plaquette is pierced by a flux of $\pi/2$. \label{fig:model}}
\end{figure}

\begin{figure}[t]
 \includegraphics[width=5cm]{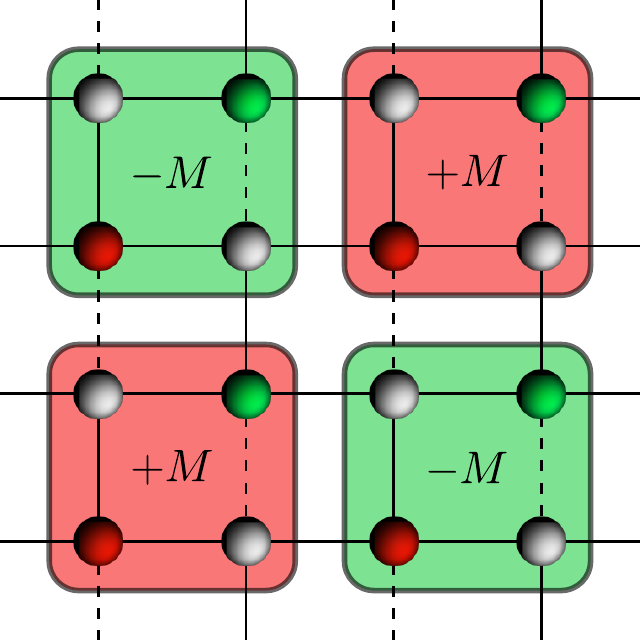}
 \caption{Additional energy offset between neighboring four-fsite plaquettes generated by the last term in Eq.~\eqref{eq:model}. Compare to Fig.~\ref{fig:model}. \label{fig:model_M}}
\end{figure}

In this section, we introduce the model under consideration. 
The Hamiltonian on the square lattice reads
\begin{align}
 H = &-J \sum_{x,y}  \{ \hat a_{x+1,y}^\dag \hat a_{x,y}^\phd e^{i[\pi/2(x+y)-\phi_0]} + \mathrm{h.c.} \nonumber \\
 & \hspace{13.5mm} + (1+f_{x,y}) \hat a_{x,y+1}^\dag \hat a_{x,y}^\phd + \mathrm{h.c.} \} \nonumber \\
 &+ \frac{\delta}{2} \sum_{x,y} [(-1)^x + (-1)^y] \hat n_{x,y} \nonumber \\
 &+ M \sum_{x,y} (-1)^{p_x+p_y} \hat n_{x,y} \label{eq:model}
\end{align}
with $\hat a_{x,y}^\dag ( \hat a_{x,y}^\phd)$ creating (annihilating) a boson on site $(x,y)$. 
We assume that the onsite interaction between particles is much larger than the hopping $J$ and work in the limit of hardcore bosons restricting the occupation number to $\hat n_{x,y} = 0,1$.
The parameters $f_{x,y}$ and $p_i$ are given by
\begin{equation}
 f_{x,y} = - \frac12 \left( \frac{\kappa}{\hbar \omega} \right)^2 \{ 1 - (-1)^{x+y}\cos(2 \phi_0) \} \label{eq:fmn},
\end{equation}
and
\begin{equation}
 p_i =
 \begin{cases}
  0, \quad \text{if} \quad i \in 4\mathbb{N} \text{ or } i \in 4\mathbb{N}+1 \\
  1, \quad \text{if} \quad i \in 4\mathbb{N}+2 \text{ or } i \in 4\mathbb{N}+3 \\
 \end{cases}
\end{equation}
Let us first focus on the terms with prefactors $J$ and $\delta$ which represent the effective Hamiltonian describing the experiment in~\cite{Aidelsburger2015}.
The hopping part describes the Hofstadter model on a square lattice with a flux of $\phi = \pi / 2$ per plaquette as shown in Fig.~\ref{fig:model} and the hopping in $y$-direction is renormalized in second-order Floquet theory by the site-dependent parameter $f_{x,y}$ given in \eqref{eq:fmn}. We chose $\kappa (\hbar \omega) = 0.58 $ as in the experiment in~\cite{Aidelsburger2015}.
At the single-particle level, the energy splits up into four bands and the term proportional to the parameter $\delta$ introduces a staggered potential along the $x$- and $y$-direction which can be used to tune the band structure from a topological to a trivial one.
A phase transition occurs at $\delta \approx 1.7$ when the Chern number of the lowest band changes from $C=1$ for $\delta \lesssim 1.7$ to $C=0$ for $\delta \gtrsim 1.7$~\cite{Aidelsburger2015}.
%
In addition to the effective Hamiltonian from \cite{Aidelsburger2015}, we introduce a term that generates an overall energy offset between neighboring unit cells in the last line of Eq.~\eqref{eq:model}. 
A schematic depiction is shown in Fig.~\ref{fig:model_M}. 

We investigate the phase diagram of Hamiltonian~\eqref{eq:model} using the DMRG algorithm~\cite{White1992,McCulloch2008,Schollwoeck2011}, on an infinite cylinder geometry~\cite{He2017}. This method allows us to variationally determine the exact ground state  for large system sizes. Most of the computations in this work are performed on a cylinder with a circumference of $L_y=8$ sites.

\section{FCI state at $\delta=M=0$ \label{sec:FCI}}
\begin{figure}[t]
 \includegraphics[width=8cm]{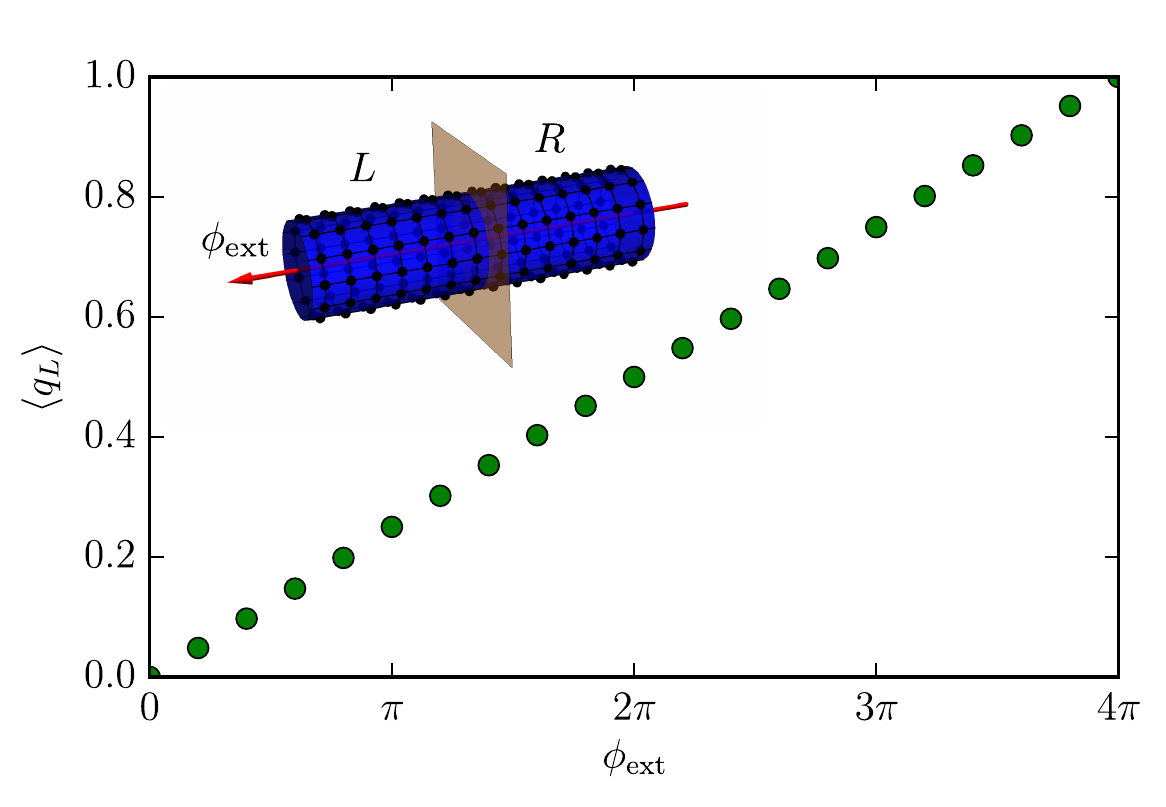}
 \caption{Charge $ \langle q_L \rangle$ of the left half of the system vs.~inserted flux for a cylinder of circumference $L_y=8$. After two flux quanta ($4 \pi$) have been inserted, one elementary charge is pumped across the cut indicating a Hall conductivity of $\sigma_{xy} = 1/2$.  \label{fig:flux}}
\end{figure}

One of the clearest signatures of the FCI state is the quantization of the Hall conductivity $\sigma_{xy}$ to fractional values. Here, we determine the $\sigma_{xy}$ by numerically conducting a Laughlin-like charge pumping experiment~\cite{Laughlin1981}. We cut the cylinder into two semi-infinite halves and monitor the charge $\langle q_L \rangle$ of the left half as we adiabatically insert flux into the cylinder. Figure~\ref{fig:flux} shows a Hall conductivity of $1/2$ in accordance with a Laughlin state at filling $\nu=1/2$~\cite{He2017,Gerster2017}.

\section{Phase diagram \label{sec:PD}}

\begin{figure*}[t]
 \includegraphics[width=\textwidth]{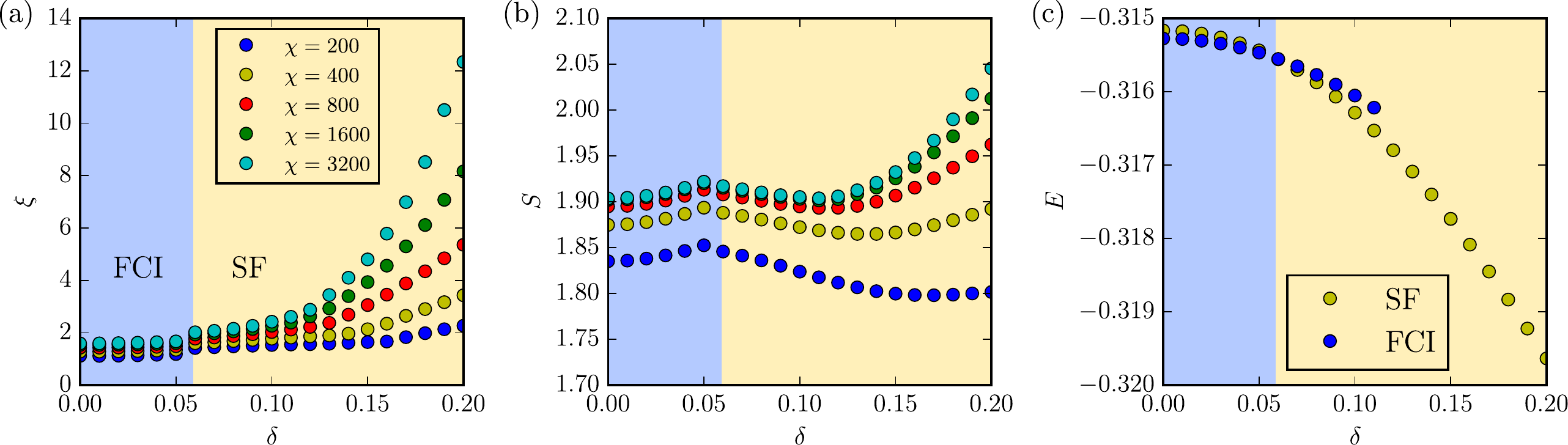}
 \caption{Correlation length $\xi$ (a) and entanglement entropy $S$ (b) as a function of $\delta$ for increasing DMRG bond dimensions $\chi$. The discontinuity in both quantities at the phase boundary of FCI and SF indicates a first-order transition. (c) Ground state energy per site around the transition between FCI and SF state. The crossing of the energy levels further corroborates the first-order nature of the transition. \label{fig:S_xi_E}}
\end{figure*}


After having established the emergence of the FCI state for $\delta=M=0$, we now investigate the phase diagram of the model \eqref{eq:model}. We first focus on $\delta > 0$, $M=0$ and then turn to the case of finite $M$.

\subsection{$\delta > 0, M = 0$}

The simplest way to tune our model into a non-topological state is increasing $\delta$. At a finite value of $\delta>0$, the band structure becomes trivial which will not permit an FCI state to occur in this model. However, the nonzero Chern number of the lowest band is merely a necessary condition for the FCI and its stability might break down even before the band structure becomes trivial. We therefore investigate the state of the system for increasing $\delta$. Let us first consider a perturbative picture of the ground state for large $\delta$. In this limit, three of the sites in the unit cell, namely the ones with energy offset $+\delta$ and $0$, are not part of the low energy subspace of the Hilbert space. The particles will prefer to occupy the sites with energy offset $-\delta$ due to the large energy penalty on the remaining three sites. For large, but finite $\delta$, we can therefore write down an effective Hamiltonian in the low energy subspace which expanded up to second order in $J$ reads
\begin{equation}
 H_{\rm low} \approx - J^2/\delta \sum_{\langle ij \rangle, i,j \in \Lambda_{-\delta}} a_i^{\dag} a_j\phd \label{eq:H_eff},
\end{equation}
where we have neglected a constant energy offset per site. Here,  $\Lambda_{-\delta}$ denotes the set of lattice sites with chemical potential $-\delta$ and the Hamiltonian describes a regular hopping model of hardcore bosons on the square lattice formed by the ``$-\delta$ sites.'' Note that the complex phases in the hopping may be gauged away since the flux through each square plaquette surrounded by the remaining sites $\Lambda_{-\delta}$ is $2 \pi$. The total filling factor of $1/8$ leads to a half-filling of the effective low-energy subspace and the ground state of this Bose-Hubbard model is well known to be superfluid, spontaneously breaking the $U(1)$ particle number conservation symmetry~\cite{Fisher1989}.

In the following, we study the stability of the FCI state in model \eqref{eq:model} identified in Sec.~\ref{sec:FCI} for increasing $\delta$ towards this superfluid state and possible competing phases.
We detect phase transitions by studying the behavior of the entanglement entropy, the correlation length or, in case of a first-order transition, the energy of the ground state~\cite{Motruk2015}. The entanglement entropy $S$ is defined as
\begin{equation}
 S = -\mathrm{Tr} \rho_L \log \rho_L,
\end{equation}
where $\rho_L = \mathrm{Tr}_R \ket{\psi}\bra{\psi}$ is the reduced density matrix of the left half $L$ of the cylinder.
In Figs.~\ref{fig:S_xi_E}(a) and \ref{fig:S_xi_E}(b), we show $S$ and the correlation length $\xi$ for increasing delta and observe that the FCI state at $\delta=0$ is stable in a finite region for $\delta > 0$.
The clear discontinuities in $\xi$ and $S$ indicate a first-order transition which is further confirmed by the behavior of $E$ depicted in Fig.~\ref{fig:S_xi_E}(b). At the transition point, we observe a crossing of the energy of the low and high $\delta$ ground states. The blue dots show the ground state energy of the FCI state while the yellow dots indicate the neighboring state. The data points at values of $\delta$ at which the respective phase is not the ground state of the system where obtained by initializing the algorithm with the state of the respective phase and converging to the local energy minimum in the Hilbert space. This energy data clearly demonstrates a transition at $\delta_c \approx 0.06$. We do not observe any sign of a phase transition for $\delta > \delta_c$ up to values of $\delta \gg J$ which represent the regime in which the system is described by the effective Hamiltonian of Eq.~\eqref{eq:H_eff}.

\subsection{$\delta > 0, M > 0$ \label{sec:M}}

Since we identified the finite-$\delta$ transition for $M=0$ to be of first order, tuning into the FCI state from the superfluid phase for large $\delta$ along a path in parameter space decreasing delta is not a favorable strategy. The likelihood for the occurrence of a second order transition might be enhanced if the state neighboring the FCI does not spontaneously break any symmetries. Motivated by this intuition, we introduce the term $\propto M$ in the Hamiltonian \eqref{eq:model} which generates an alternating additional chemical potential on every four-site unit cell of the original Hamiltonian of the experiment. 

\begin{figure}[b]
 \includegraphics[width=\columnwidth]{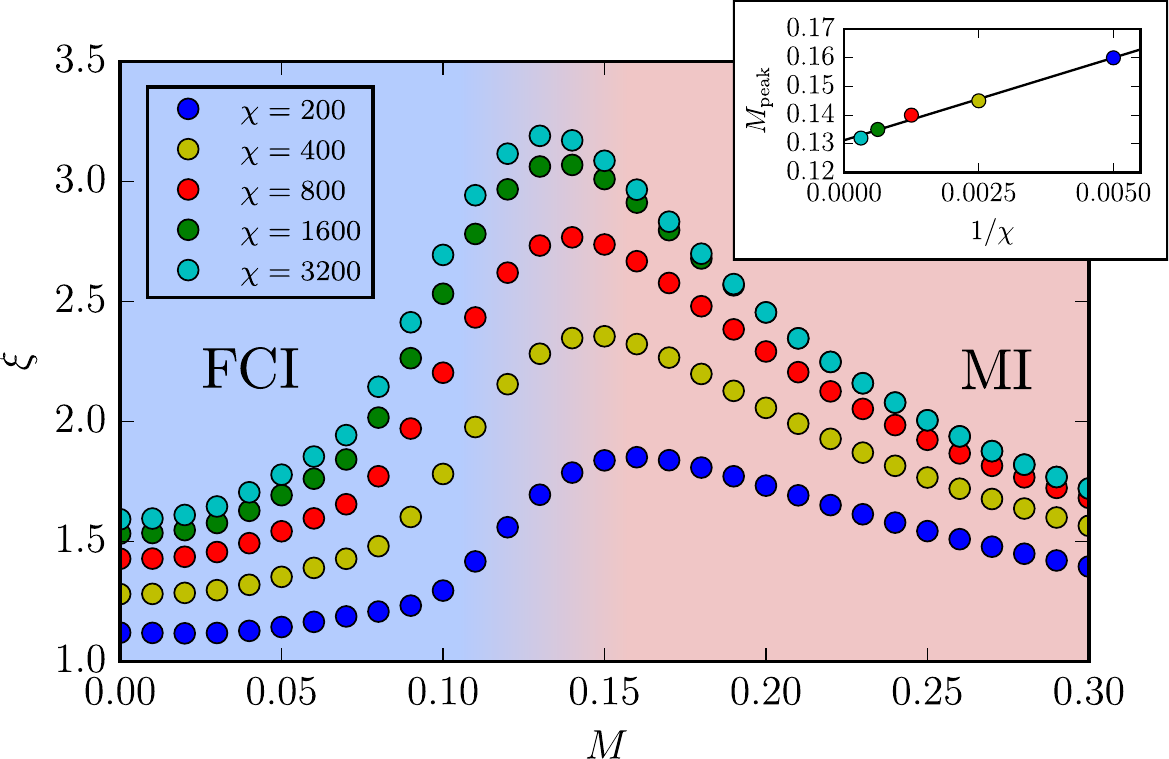}
 \caption{Correlation length $\xi$ for $\delta = 0$ and finite $M$. The peak in $\xi$ and its increase towards higher bond dimensions suggests a continuous transition between the large $M$ and the FCI state. We also checked the Hall conductivity which is quantized to $\sigma_{xy}=1$ below the value of $M$ at which the system is entering the critical regime, e.g., for $M < 0.1$ for $\chi = 400$. Inset: Scaling of $M_{\rm peak}$, the position of the peak in $\xi$ vs.\ $1 / \chi$ to determine the transition point for $\chi \rightarrow \infty$.  \label{fig:corr_M}}
\end{figure}

Let us again consider the limit in which $\delta, M \gg J$. By adding the above term to the Hamiltonian, we obtain one site with energy offset $-\delta-M$ per each eight sites in the system (the green sites in the green squares in Fig.~\ref{fig:model_M}). The number of these sites is hence exactly the number of particles and all other sites are at least $\delta$ or $2M$ higher in energy, whichever of these numbers is smaller. In the limit of $\delta, M \rightarrow \infty$, these sites will therefore each be occupied by exactly one particle. Upon decreasing $\delta$, the particles will be allowed to hop around a four-site plaquette, but will still be localized and form a Mott insulating (MI) phase. This phase does not spontaneously break any symmetry and therefore constitutes a phase which might feature a continuous transition into the FCI state.

\begin{figure}[b]
 \includegraphics[width=8cm]{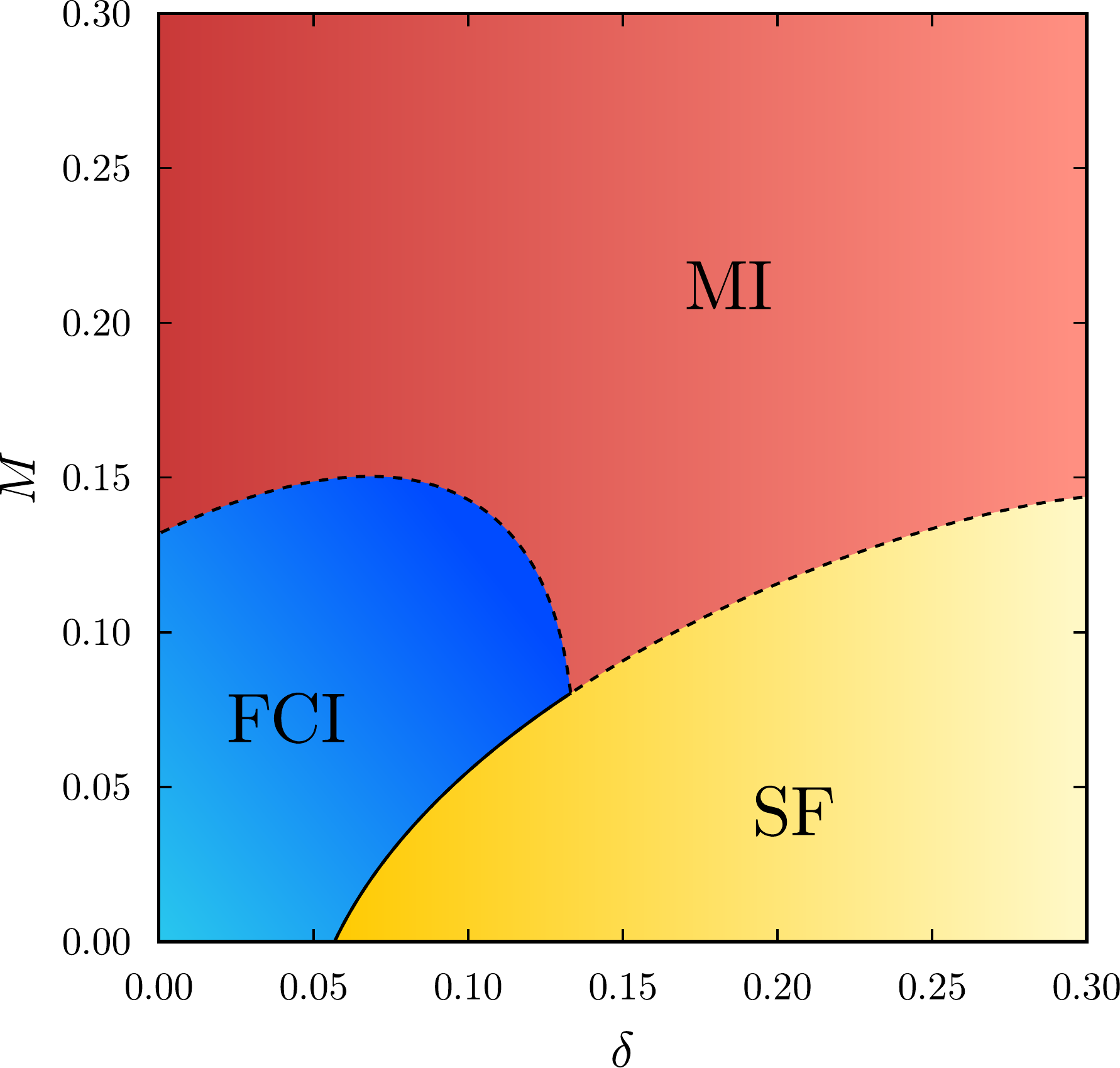}
 \caption{Phase diagram in the $\delta$-$M$ plane. The FCI state is stable in a finite region in the phase diagram. The first-order transition between FCI and superfluid is depicted by a solid line, dashed lines denote continuous transitions. \label{fig:pd}}
\end{figure}

In Fig.~\ref{fig:corr_M}, we plot the correlation length $\xi$ of the ground state for $\delta = 0$ and finite $M$. We observe a peak in $\xi$ at $M \approx 0.13$, but no discontinuity as in the transition between SF and FCI. The behavior in $\xi$ indeed points to a continuous transition between the two phases. The peak in $\xi$ signals a critical point. In the region around the point, the state cannot be faithfully represented with finite DMRG bond dimension $\chi$ anymore and the correlation length still grows with increasing $\chi$. This entrance into the critical region is as well reflected in the Hall conductivity which is not quantized anymore when approaching the transition from the FCI side~\cite{Gerster2017}. 

\begin{figure*}[t]
 \includegraphics[width=\textwidth]{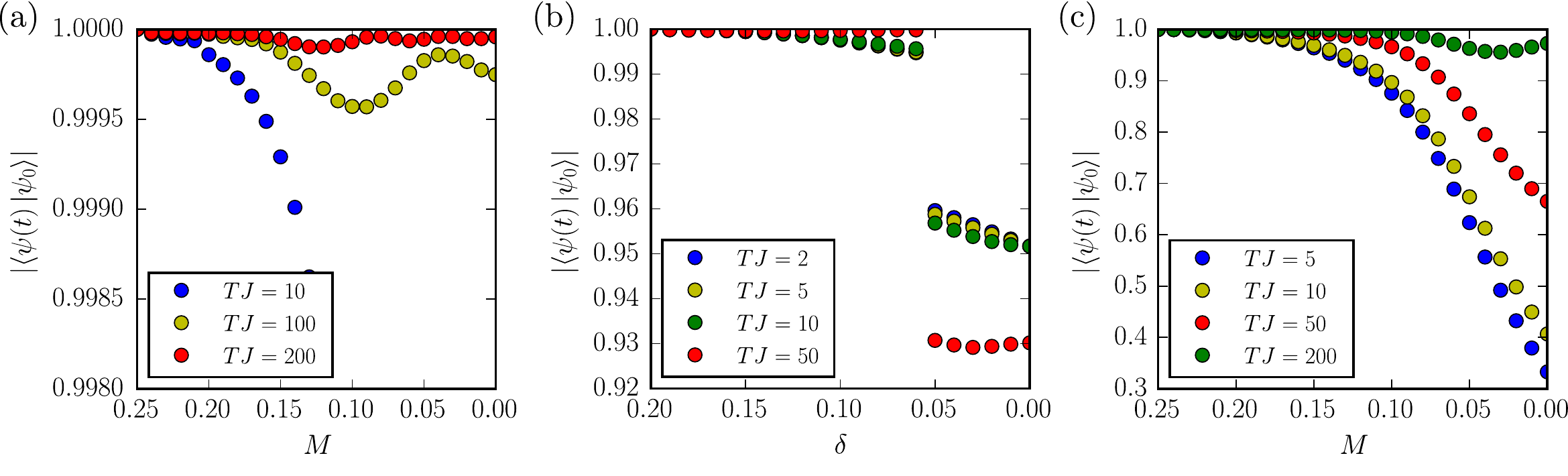}
 \caption{Overlap (per site) of the time-evolved state $\ket{\psi(t)}$ with the actual ground state $\ket{\psi_0}$ at the respective parameter when tuning the Hamiltonian. (a) Fixing $\delta=0$ and tuning $M$ from $M_0 = 0.25$ to $0$ for an infinite cylinder of circumference $L_y = 8$. Note that $M$ decreases to the right which means time increases. The overlap per site $\left\langle \psi(t) \right| \left. \psi_0 \right\rangle$ increases for slower evolution and reaches $\sim 0.9999$ for $TJ=200$ at $M=0$. (b) Tuning the parameter $\delta$ from $\delta_0 = 0.2$ to $0$ with $M=0$ through the first-order transition on an infinite cylinder. The overlap per site sharply decreases when crossing the phase transition and cannot be increased by a slower evolution. (c) Time evolution of the state on a finite square geometry of $L_x \times L_y = 8 \times 8$ sites for the same path in parameter space as in (a). The overlap per site of the final state is significantly lower ($\sim 0.973$ for $TJ = 200$) than in the case of the infinite system, but approaches unity for longer evolution times. In all calculations, the bond dimension $\chi$ is chosen such that the value of the overlap has converged in $\chi$.
 \label{fig:overlap}}
\end{figure*}
%

We map out the full phase diagram of the model \eqref{eq:model} in the $\delta$-$M$ parameter plane shown in Fig.~\ref{fig:pd}. The FCI region proves to be stable in a finite parameter region and is separated by a second order transition from the Mott insulating (MI) state without a spontaneously broken symmetry. We observe a similar behavior of the correlation length as in Fig.~\ref{fig:corr_M} at the transition between FCI and MI for finite $\delta$. The transition into the superfluid with spontaneously broken $U(1)$ symmetry is of first order. 

While the simulations presented in this work were conducted for hardcore bosons, an experimental realization of the model will have finite interactions.  When releasing the hardcore constraint, a sufficiently strong on-site repulsion is required to stabilize the FCI phase. If the interactions are strong enough, the transition between MI and FCI remains continuous as indicated by numerical simulations (not shown).

\section{Adiabatic tuning \label{sec:time}}

The continuous transition in the model from a trivial insulating to an FCI state harbors the potential for an adiabatic preparation of the topological state. To further corroborate this possibility, we simulate the preparation scheme along the line $\delta = 0$ in the phase diagram by means of time-dependent DMRG calculations. We use the method introduced by Zaletel \textit{et al.}~\cite{Zaletel2015} which allows us to treat the effective long-range interactions that occur in a 2D DMRG setup~\cite{Stoudenmire2012}. 

\subsection{Infinite cylinder}

We first study the system in the infinite cylinder geometry as before and time-evolve from an initial state $\ket{\psi_I}$ to a final state $\ket{\psi_F}$ under Hamiltonian~\eqref{eq:model}
with the time-dependent parameter
\begin{equation}
 M(t) = M_0 (1-t/T), \quad 0 \leq t \leq T.
\end{equation}
We start the evolution with the ground state at $M_0 = 0.25$ which is well separated from the transition into the MI phase. In Fig.~\ref{fig:overlap}(a), we plot the overlap (per site) of the time-evolved state with the actual ground state at the respective parameter value. We observe a high overlap which steadily increases for longer ramp times. This indicates that the adiabatic evolution may be performed successfully for sufficiently long times.

%

In order to demonstrate the importance of crossing a continuous rather than a first-order phase transition in the adiabatic preparation, we also calculate the time evolution for a path from the superfluid phase to the FCI phase traversing the first-order transition. We therefore set $M=0$ and evolve $\delta$ from $\delta_0 = 0.2$ to $0$ as
\begin{equation}
 \delta(t) = \delta_0 (1-t/T).
\end{equation}
The results are presented in Fig.~\ref{fig:overlap}(b). We clearly observe that the adiabatic preparation is not successful along this path in the phase diagram. The overlap significantly decreases at the transition point and it is not possible to remedy this behavior by choosing a longer ramp time. Instead, the overlap even decreases for longer times. The jump in the overlap may be explained by the nature of the first-order transition. As depicted in Fig.~\ref{fig:S_xi_E}(c), the energy levels of SF and FCI cross at the transition. When (quasi-)adiabatically tuning the Hamiltonian parameters from the SF side across the transition, the system then tries to follow the (metastable) SF state and does not evolve into a state with high overlap to the actual FCI ground state. This scenario is further confirmed by the larger decrease in overlap for longer ramp times, indicating that the state follows the initial SF state better for slower evolution. Since we show the overlap per site, the total overlap for a system of $N$ sites would amount to $\sim 0.9^N$ so that the final state would be far from the desired FCI state.

\subsection{Finite square geometry}


We also simulate the preparation protocol on a finite square geometry of $8 \times 8$ sites in order to be closer to a potential experimental setup with edges. The results are depicted in Fig.~\ref{fig:overlap}(c). The overlap of the final state with the actual ground state is much less than in the infinite cylinder geometry. This behavior can be explained by the presence of gapless --or low energy for finite system size-- edge modes and is consistent with what has been observed in Ref.~\cite{He2017}. However, our results indicate that the FCI may still be prepared for sufficiently long ramp times.

\section{Conclusion}

We have investigated the Harper-Hofstadter Hamiltonian describing the cold atom experiment by Aidelsburger et al.~\cite{Aidelsburger2015} at half filling of the lowest band for hardcore bosons. For the pure hopping Hamiltonian, we find a topologically ordered fractional Chern insulator state which remains stable for a finite staggered chemical potential term with parameter $\delta$ which is present in the original model. The increasing of this potential tunes the band structure from topological to trivial. However, the FCI state proves to be unstable at a potential of $\delta_c \approx 0.06$ much below the value when the underlying band structure becomes non-topological $(\delta \approx 1.7)$ and the system undergoes a first-order phase transition into a superfluid state. 
In addition to the staggered potential from the original experiment, we introduce another periodic chemical potential term with parameter $M$ which tunes the system into a trivial Mott insulator. We map out the phase diagram as a function of $\delta$ and $M$ and find that the transition from the Mott insulator into the FCI state is of second order providing a path along which the FCI state may be adiabatically prepared.

To demonstrate this adiabatic preparation scheme, we explicitly simulate the time evolution of a trivial initial state to the FCI state when slowly tuning the parameters of the Hamiltonian. On an infinite cylinder, we find an overlap per site of $\sim 0.9999$ of the time-evolved state with the FCI ground state for a ramp time of $T=200/J$ which may be further improved by increasing the time. When considering a finite square geometry, the final overlap reduces to $\sim 0.973$ for otherwise equal parameters, but a sufficiently slow tuning in an experiment should still guarantee a topologically ordered final state. Our protocol requires only a rather small modification of the one already implemented experimentally.

\section*{Acknowledgments}

We thank Adolfo Grushin, Michael Kolodrubetz and Gunnar M\"oller for comments on the manuscript. JM acknowledges funding by TIMES at Lawrence Berkeley National Laboratory supported by the U.S. Department of Energy, Office of Basic Energy Sciences, Division of Materials Sciences and Engineering, under Contract No.\ DE-AC02-76SF00515. This work was supported in parts by the German Research Foundation (DFG) via the
Collaborative Research Center SFB 1143 and Research Unit FOR 1807 through grants no.\ PO 1370/2-1.

\end{document}